\documentclass[11pt,reqno]{amsart}

\usepackage{amsmath}
\usepackage{amssymb}

\newtheorem{Theorem}{Theorem}[section]

\newtheorem{Corollary}[Theorem]{Corollary}

\theoremstyle{definition}
\newtheorem{Remark}[Theorem]{Remark}
\newtheorem{Example}[Theorem]{Example}

\numberwithin{equation}{section}

\begin{document}

\title[Ruijsenaars-Schneider hierarchy and bispectral 
operators]{Rational Ruijsenaars-Schneider hierarchy and bispectral 
difference operators}

\author{Plamen Iliev}
\address{School of Mathematics, Georgia Institute of Technology, 
Atlanta, GA 30332--0160, USA}
\email{iliev@math.gatech.edu}

\date{November 14, 2006}

\newcommand{\seref}[1]{Section \ref{#1}}
\newcommand{\thref}[1]{Theorem \ref{#1}}
\newcommand{\leref}[1]{Lemma \ref{#1}}
\newcommand{\prref}[1]{Proposition \ref{#1}}
\newcommand{\exref}[1]{Example \ref{#1}}

\newcommand{\cC}{{\mathcal C}}
\newcommand{\cA}{{\mathcal A}}
\newcommand{\cB}{{\mathcal B}}
\newcommand{\cL}{{\mathcal L}}

\newcommand{\De}{\Delta}

\newcommand{\C}{\mathbb C}
\newcommand{\Z}{\mathbb Z}
\newcommand{\N}{\mathbb N}
\newcommand{\tr}{\mathrm{tr}}
\newcommand{\diag}{\mathrm{diag}}
\newcommand{\Spec}{\mathrm{Spec}}
\newcommand{\rank}{\mathrm{rank}}
\newcommand{\Grad}{\mathrm{Gr}^{\mathrm{ad}}}

\newcommand{\pd}{\partial}
\newcommand{\Mh}{{\hat{M}}}
\newcommand{\Yh}{{\hat{Y}}}

\newcommand{\Xt}{{\tilde{X}}}
\newcommand{\Yt}{{\tilde{Y}}}
\newcommand{\xt}{{\tilde{x}}}
\newcommand{\yt}{{\tilde{y}}}
\newcommand{\taut}{{\tilde{\tau}}}

\newcommand{\expp}{\exp\left(\sum_{i=1}^{\infty}t_iz^i\right)}
\newcommand{\expm}{\exp\left(-\sum_{i=1}^{\infty}t_iz^i \right)}
\newcommand{\Expp}{{\mathrm{Exp}}(n;t,z)}
\newcommand{\Expm}{{\mathrm{Exp}}^{-1}(n;t,z)}

\newcommand{\vw}{{\vec{w}}}
\newcommand{\ve}{{\vec{e}}}


\begin{abstract} 
We show that a monic polynomial in a discrete variable $n$, with coefficients 
depending on time variables $t_1, t_2,\dots$ is a $\tau$-function 
for the discrete Kadomtsev-Petviashvili hierarchy if and only if the motion 
of its zeros is governed by a hierarchy of Ruijsenaars-Schneider 
systems. These $\tau$-functions were considered in \cite{HI}, where it was 
proved that they parametrize rank one solutions to a 
difference-differential version of the bispectral problem.
\end{abstract}

\maketitle
\section{Introduction}

In \cite{AMcKM}, Airault, McKean and Moser discovered a mysterious connection 
between equations of KdV type and the Calogero-Moser system.  
They showed that the motion of the poles of a rational solution to the KdV or 
Boussinesq equation that vanishes at infinity is described by the 
Calogero-Moser system \cite{C}, with some constraint on the configuration of 
poles. Krichever \cite{K} observed that the poles of the 
rational solutions to the KP equation that 
vanish at $x=\infty$, move according to the Calogero-Moser system with no 
constraint. Shiota \cite{Sh} extended this phenomenon to the whole 
KP hierarchy, which combined with the work of Adler \cite{A} led to a simple  
explicit formula for the $\tau$-function.

A surprising link to the above theory was observed by 
Duistermaat and Gr\"unbaum \cite{DG} in connection with a problem in limited 
angle tomography \cite{Gr}, known as the bispectral problem.  As originally 
formulated, this problem asks for which ordinary
differential operators $L(x,d/dx)$ there exists a family of eigenfunctions 
$\Psi(x,z)$ that are also eigenfunctions for another differential operator 
$B(z,d/dz)$ in the ``spectral parameter'' $z$. In the case when the operator 
$L(x,d/dx)$ belongs to a rank one commutative ring of differential operators
(i.e. $L$ commutes with an operator of odd order), the solution of the 
bispectral problem (up to translations and rescalings
of $x$ and $z$) are precisely the operators which can be obtained by finitely 
many rational Darboux transformations from $L_0=d^2/dx^2$. This combined with 
work of Adler and Moser \cite{AM} shows that the rank one solutions of the 
bispectral problem are exactly the rational solutions discovered in 
\cite{AMcKM}. 
Wilson \cite{W1} proposed to extend the problem to commutative 
rings of differential operators. Such a ring is called bispectral when 
there is a joint eigenfunction of the operators in the ring that is also a 
joint eigenfunction of a ring of differential operators in the spectral 
variable. An important invariant of such a 
ring is its rank, meaning the dimension of the common space of eigenfunctions 
to the operators belonging to the ring. He proved that the bispectral maximal 
rank one commutative rings of differential operators are parametrized by a 
sub-Grassmannian $\Grad$ of Sato's Grassmannian \cite{SS}, which corresponds 
to the rational solutions of the KP equation studied by Krichever \cite{K}. 
Moreover, in a subsequent paper \cite{W2}, Wilson gave a beautiful explanation 
of the bispectral property based on the connection with Calogero-Moser systems 
and their geometric description \cite{KKS}. He also deepened the mystery by 
showing that the correspondence between the Calogero-Moser and the 
KP systems extends even to the locus where the particles collide with each 
other. For a very nice characterization 
of the Grassmannian $\Grad$ in terms of representation theory see the recent 
work of Horozov \cite{Hor}. For an intriguing connection of the above 
theory to noncommutative geometry see \cite{BW}.

In \cite{HI}, jointly with Luc Haine, we constructed rank one commutative 
rings of difference operators in a discrete variable $n\in\Z$, 
corresponding to a flag of nested subspaces, 
each of which belongs to $\Grad$. We showed that the common eigenfunction 
of the operators in the ring is also the common eigenfunction of a 
maximal rank one commutative ring of differential operators in the 
spectral variable, i.e. they provide rank one solutions to a 
difference-differential version of the bispectral problem. The corresponding 
$\tau$-functions $\tau(n;t)$ are polynomials in $n$ and give rational 
solutions of the discrete KP hierarchy. 

In the present paper, we investigate 
the motion of the zeros of polynomial (in $n$) $\tau$-functions of the 
discrete KP hierarchy. We show that a monic polynomial in $n$ of degree $N$ is 
a $\tau$-function for the discrete KP hierarchy if and only if the motion of 
its roots $\{x_i\}_{i=1}^N$ is governed by a hierarchy of 
Ruijsenaars-Schneider systems. We restrict out attention to the generic 
situation when the roots satisfy the constraints $x_i-x_j\notin\{0,1\}$ for 
$i\neq j$. This condition means that the rational solution to the simplest 
zero curvature (Zakharov-Shabat) equation for the discrete KP hierarchy has 
$2N$ distinct poles. It is a challenging problem to investigate the more 
general case allowing collisions of the poles.

The paper can be thought of as a discrete analog of Shiota's paper \cite{Sh}. 
In particular, from the proof, we can easily write an explicit formula for 
the $\tau$-functions in terms of 
the Ruijsenaars-Schneider hierarchy, which implies that they parametrize 
the rank one solutions of the difference-differential version of 
the bispectral problem constructed in \cite{HI}.  

We note that there is a related work of van Diejen \cite{vD} in the case of 
second-order difference operators, where the dynamics of the zeros of the 
solitonic Baker-Akhiezer function in the 
spectral variable $z$ is studied. For soliton solutions of KP and 2D Toda 
equations, see Ruijsenaars \cite{R} and van Diejen-Puschmann \cite{vDP},
and for elliptic generalizations see 
Krichever-Zabrodin \cite{KZ}. For a $q$-deformation of the KP hierarchy and 
connections with the bispectral problem see \cite{I2}.

The paper is organized as follows. In the next section, we briefly introduce 
the necessary ingredients of the discrete KP hierarchy. The approach follows 
closely \cite{HI}, which leads to $\tau$-functions that differ by an 
exponential factor from the ones constructed in \cite{AvM,UT}. In 
\seref{se3} we formulate the main result of the paper and its connection to 
the bispectral problem. For a very nice account on the 
difference-differential version of the bispectral problem and its 
relations to orthogonal polynomials and the Toda lattice see \cite{H}.
\seref{se4} is devoted to the proof of the main result.

\section{The discrete KP hierarchy and $\tau$-function}         \label{se2}

We denote by $\De$ and $\nabla$ the customary forward and backward difference 
operators acting on functions of a discrete variable $n\in\Z$ by
\begin{equation*}
\De f(n) = f(n+1)-f(n) \text{ and }\nabla f(n)=f(n)-f(n-1).
\end{equation*}
The formal adjoint to $\De$ is $\De^*=-\nabla$. If we define
\begin{equation*}
\De^j\cdot f(n)=\sum_{i=0}^{\infty}\binom{j}{i}(\De^if)(n+j-i)\De^{j-i},
\qquad \text{ for all }j\in\Z,
\end{equation*}
we obtain an associative ring of formal pseudo-difference operators
\begin{equation*}
R\{\De\}=\Bigg\{X=\sum_{j=-\infty}^da_j(n)\De^j\Bigg\}.
\end{equation*}
We denote by $X_+=\sum_{j=0}^da_j(n)\De^j$ the positive difference 
part of $X$ and by $X_-=\sum_{j=-\infty}^{-1}a_j(n)\De^j$, the Volterra 
part of $X$. 

The discrete Kadomtsev-Petviashvili hierarchy (in short KP) 
is the family of evolution equations in 
infinitely many time variables $t=(t_1,t_2,t_3,\dotsc)$ given by 
the Lax equations
\begin{equation}                                                \label{2.1}
\frac{\pd L}{\pd t_i}=[(L^i)_+,L],
\end{equation}
where $L$ is a general formal pseudo-difference operator of the form
\begin{equation*}
L=\De+\sum_{j=0}^{\infty}a_j(n)\De^{-j}.
\end{equation*}
A $\tau$-function for the hierarchy \eqref{2.1} can be defined as follows. 
First, we define a wave operator 
\begin{equation*}
W(n;t)=1+\sum_{j=1}^{\infty}w_j(n;t)\De^{-j},
\end{equation*}
which conjugates $L$ to $\De$, that is
\begin{equation}                                                \label{2.2}
L=W\De W^{-1}.
\end{equation}
The vector fields \eqref{2.1} can be extended by 
\begin{equation}                                                \label{2.3}
\frac{\pd W}{\pd t_k}=-(L^k)_-W.
\end{equation}
For simplicity, we denote by $\Expp$ the exponential function 
\begin{equation*}
\Expp= (1+z)^n\expp.
\end{equation*}
The wave function $w(n;t,z)$ and the adjoint wave function 
$w^*(n;t,z)$ of the discrete KP hierarchy \eqref{2.1} are defined by
\begin{subequations}\label{2.4}
\begin{align}
w(n;t,z)&=W(n;t)\Expp\nonumber\\
&=\left(1+\frac{w_1(n;t)}{z}+\frac{w_2(n;t)}{z^2}+\dotsb\right)\Expp
                                                                \label{2.4a}\\
\intertext{and}
w^*(n;t,z)&=\left(W^{-1}(n-1;t)\right)^*\Expm\nonumber\\
&=\left(1+\frac{w^*_1(n;t)}{z}+\frac{w^*_2(n;t)}{z^2}+\dotsb\right)\Expm.
                                                                \label{2.4b}
\end{align}
\end{subequations}
The functions $w(n;t,z)$ and $w^*(n;t,z)$ can be written in terms of a 
$\tau$-function as follows
\begin{subequations}\label{2.5}
\begin{align}
&w(n;t,z)=\frac{\tau(n;t-[z^{-1}])}{\tau(n;t)}\Expp,            \label{2.5a}\\
\intertext{and}
&w^*(n;t,z)=\frac{\tau(n;t+[z^{-1}])}{\tau(n;t)}\Expm,          \label{2.5b}
\end{align}
\end{subequations}
where $[z]=(z,z^2/2,z^3/3,\dotsc)$. We refer the reader to \cite{HI} for 
more details and proofs of the above construction. 

\begin{Remark}\label{re2.1} It is well known (see for instance 
\cite[Proposition 5.1.4, p.~75]{Di}) that the Lax equations 
\eqref{2.1} imply the zero curvature (Zakharov-Shabat) equations
\begin{equation}\label{2.6}
\frac{\pd (L^k)_+}{\pd t_m}-\frac{\pd (L^m)_+}{\pd t_k}=[(L^m)_+,(L^k)_+],
\end{equation}
where $k,m\in\N$. In the differential case, the first flow $t_1$ corresponds 
to a translation in the spatial variable and therefore the simplest 
(nontrivial) zero curvature equation can be obtained for $m=2$ and $k=3$. This 
leads to the KP equation, which gave the name of the whole hierarchy. In the 
discrete case, the first flow is no longer trivial and the simplest zero 
curvature equation will correspond to the choice $m=2$ and $k=1$. In the 
example below we carry out this computation explicitly, which leads to a 
nonlinear partial differential-difference equation for the function $a_0(n;t)$.
\end{Remark}

\begin{Example}\label{ex2.2} Let us take $m=2$ and $k=1$ in \eqref{2.6}. 
Clearly, $(L)_+=\De+a_0(n;t)$ and a short computation shows that 
$$(L^2)_+=\De^2+\left(a_0(n;t)+a_0(n+1,t)\right)\De+
a_0^2(n;t)+a_1(n,t)+a_1(n+1,t).$$
Next, we see that 
\begin{equation*}
[(L^2)_+,(L)_+]=\left(\De^2 a_0(n,t)-\De\left(a_1(n;t)+a_1(n+1;t)\right)\right)
\left(\De+1\right).
\end{equation*}
Using the relations above and comparing the coefficients of $\De^i$ for 
$i=0,1$ in \eqref{2.6} with $m=2$ and $k=1$ we get the system
\begin{align*}
&-\frac{\pd \left(a_0(n;t)+a_0(n+1;t)\right)}{\pd t_1}\\
&\qquad\qquad=\De^2a_0(n;t)-\De\left(a_1(n;t)+a_1(n+1;t)\right)\\
&\frac{\pd a_0(n;t)}{\pd t_2}
-\frac{\pd \left(a_0^2(n;t)+a_1(n;t)+a_1(n+1;t)\right)}{\pd t_1}\\ 
&\qquad\qquad=\De^2a_0(n;t)-\De\left(a_1(n;t)+a_1(n+1;t)\right).
\end{align*}
Eliminating $a_1(n;t)$ we obtain the following 
equation for $a_0=a_0(n;t)$
\begin{equation}\label{2.7}
\frac{\pd }{\pd t_2}\De a_0=\frac{\pd }{\pd t_1}\left(\De a_0^2-2\De a_0\right)
+\frac{\pd^2 }{\pd t_1^2}\left(\De a_0+2a_0\right).
\end{equation}
\end{Example}

\section{Polynomial $\tau$-functions and the dynamics of their zeros}
                                                                 \label{se3}

The main result of the paper is the following theorem.

\begin{Theorem}\label{th3.1}
Let $x_{1}(t),x_{2}(t),\dots, x_{N}(t)$ be smooth functions of 
$t=(t_1,t_2,\dots)$ such that $x_i(t)-x_j(t)\notin\{0,1\}$ for $i\neq j$ 
and $\pd x_i(t)/\pd t_1\neq 0$ in a neighborhood of $t=0$. 
Let us define functions $y_1(t),y_2(t),\dots,y_N(t)$ by the following 
relation
\begin{equation}                                                \label{3.1}
e^{-y_i(t)}=-\frac{\pd x_i(t)}{\pd t_1}
\prod_{\begin{subarray}{c}s=1 \\s\neq i\end{subarray}}^N
\frac{x_i(t)-x_s(t)}{x_i(t)-x_s(t)+1}.
\end{equation}
Then the following conditions are equivalent.
\begin{itemize}
\item[(i)] The function 
\begin{equation}                                                \label{3.2}
\tau(n;t)=\prod_{i=1}^N(n-x_i(t)),
\end{equation}
is a $\tau$-function for the discrete KP hierarchy \eqref{2.1}. 
\item[(ii)] The motion of $\{x_i(t),y_i(t)\}_{i=1}^N$ is governed by 
the Ruijsenaars-Schneider hierarchy of Hamiltonian systems
\begin{equation}                                                \label{3.3}
\frac{\partial }{\partial t_k}
\left(\begin{array}{c} x_i \\y_i\end{array}\right) = (-1)^k
\left(\begin{array}{c}
\partial H_k /\partial y_i \\-\partial H_k /\partial x_i
\end{array} 
\right),\quad k=1,2,\dots,
\end{equation}
where $H_k=\tr(Y^{k})$, and $Y$ is an $N\times N$ matrix with entries
\begin{equation}                                                \label{3.4}
Y_{ij}=\delta_{i,j}+\frac{e^{-y_i}}{x_i-x_j-1}
\prod_{\begin{subarray}{c}s=1 \\s\neq i\end{subarray}}^N
\frac{x_i-x_s+1}{x_i-x_s}.
\end{equation}
\end{itemize}
\end{Theorem}

\begin{Remark}\label{re3.2} Although the Hamiltonian system above differs 
slightly from the standard rational Ruijsenaars-Schneider system, one can 
easily connect the two. Indeed, the rational Ruijsenaars-Schneider model is 
a dynamical system, whose equations of motion can be written in 
the following form
\begin{equation}\label{3.5}
\ddot{q}_j=2\sum_{\begin{subarray}{c}k=1 \\k\neq j\end{subarray}}^N
\dot{q}_j\dot{q}_k\frac{\gamma^2}{\left(\gamma^2+(q_j-q_k)^2\right)(q_j-q_k)},
\quad j=1,2,\dots,N,
\end{equation}
see \cite[formulas (B22)-(B23), p.~402]{RS}. If we now put 
$q_j=i\gamma x_j$, then \eqref{3.5} gives precisely the dynamical 
system \eqref{3.3} for the first flow $\pd/\pd t_1$, see equation \eqref{4.7}.
\end{Remark}

\begin{Remark} \label{re3.3} The solutions of the KP equation 
$$\frac{3}{4}u_{yy}=\left\{u_t-\frac{1}{4}(u_{xxx}+6uu_x)\right\}_x$$
that are rational in $x$ and vanish as $x\rightarrow\infty$ have the form 
$$u(x,y,t)=-2\sum_{j=1}^N\frac{1}{(x-x_j(y,t))^2}.$$
When all $x_j$ are distinct their motion is governed by the Calogero-Moser 
system, as shown in \cite{K}. The case discussed in the present paper is 
similar in the following sense: the solutions described in \thref{th3.1} 
provide rational solutions of equation \eqref{2.7}, which vanish as 
$n\rightarrow\infty$. These solutions have simple poles (as functions of $n$)
at the points $\{x_j,x_j-1\}_{j=1}^N$, see formula \eqref{4.3}. The condition 
$x_i-x_j\notin\{0,1\}$ for $i\neq j$ simply means that all these poles are 
distinct.
\end{Remark}

As a consequence of the proof of \thref{th3.1} we also obtain an explicit 
formula for the $\tau$-function in terms of $\{x_i,y_i\}_{i=1}^N$ at 
$t_1=t_2=\cdots=0$. Let us denote by $X$ the diagonal matrix with entries 
$x_i(t)$, i.e. 
\begin{equation}\label{3.5a}
X=\diag(x_1(t),x_2(t),\dots,x_N(t)).
\end{equation}
\begin{Corollary}\label{co3.4}
Let $X^0$ and $Y^0$ be the matrices $X$ and $Y$, defined by \eqref{3.5a} and 
\eqref{3.4} at $t_1=t_2=\cdots=0$. Then the $\tau$-function in equation 
\eqref{3.2} can be computed from the following formula
\begin{equation}\label{3.6a}
\tau(n;t)
=\det\Big(nI-X^0+\sum_{j=1}^{\infty}jt_j(I-Y^0)(-Y^0)^{j-1}\Big),
\end{equation}
where $I$ is the identity $N\times N$ matrix.
\end{Corollary}
From formula \eqref{3.6a} it is easy to see that 
$\tau(n;t)=\tau(0;t_1+n,t_2-n/2,t_3+n/3,\dots)$, where $\tau(0;t)$ is 
a $\tau$-function for the (continuous) KP hierarchy, corresponding to 
a plane in Wilson's adelic Grassmannian. Thus, the results in \cite{HI} 
imply that the functions $\tau(n;t)$ described in \thref{th3.1} parametrize 
rank-one solutions to a difference-differential version of the bispectral 
problem. More precisely, there exist a rank-one commutative ring $\cA$ of 
difference operators in the variable $n$, and a rank-one commutative ring 
$\cA'$ of differential operators in $z$, such that 
\begin{align*}
\cL w(n;t,z)& = f_{\cL}(z)w(n;t,z), \qquad\forall \cL\in\cA\\
\cB w(n;t,z)& = g_{\cB}(n)w(n;t,z), \qquad\forall \cB\in\cA'
\end{align*}
where $f_{\cL}(z)$ and $g_{\cB}(n)$ are functions of $z$ and $n$, 
respectively, and $w(n;t,z)$ is the wave function defined by \eqref{2.5a}.

\section{Proof of Theorem 3.1}                          \label{se4}
The strategy of the proof is as follows. For the implication 
(i)$\Rightarrow$(ii), we investigate the motion 
of the poles with respect to the first flow $\pd/\pd t_1$ and we write 
the corresponding dynamical system in an appropriate Lax form. This 
represents a discrete analog of some of the results in \cite{K,M}, except that 
in the continuous case the first nontrivial flow is $\pd/\pd t_2$. Next, we 
adapt the approach in \cite{Sh} to establish the Hamiltonian equations for 
the higher flows $\pd/\pd t_k$, $k\geq 2$. The opposite direction can be 
deduced by using the connection between polynomial (in $t_1$) $\tau$-functions 
of the KP hierarchy and polynomial (in $n$) $\tau$-functions of the discrete 
KP hierarchy \cite{HI}.

Let us start with the implication (i)$\Rightarrow$(ii).
From equations \eqref{2.4}, \eqref{2.5} and \eqref{3.2} it is clear that we 
can write $w_k(n;t)$ and $w^*_k(n;t)$ as 
\begin{subequations}\label{4.1}
\begin{align}                                                
w_k(n;t)&=\sum_{i=1}^N \frac{w_{k,i}(t)}{n-x_i(t)}      \label{4.1a} \\
w^*_k(n;t)&=\sum_{i=1}^N \frac{w^*_{k,i}(t)}{n-x_i(t)}. \label{4.1b} 
\end{align}
\end{subequations}
In particular, for $k=1$ we see that $w_{1,i}(t)=\pd x_i(t)/\pd t_1$
and $w^*_{1,i}(t)=-\pd x_i(t)/\pd t_1$.
From \eqref{2.2}, \eqref{2.3} and \eqref{2.4a} it follows that 
\begin{equation}                                                \label{4.2}
\frac{\pd w(n;t,z)}{\pd t_1}=(\De+a_0(n;t))w(n;t,z).
\end{equation}
Writing \eqref{2.2} as $LW=W\De$ and comparing the coefficients of $\De^0$ on
both sides gives
\begin{align}
a_0(n;t) & =-w_1(n+1;t)+w_1(n;t)\nonumber \\
& = \sum_{i=1}^N \frac{1}{(n-x_i(t))(n+1-x_i(t))} 
\frac{\pd x_i(t)}{\pd t_1}\label{4.3}
\end{align}
where in the last equality we used \eqref{4.1a} for $k=1$. Plugging the last 
formula for $a_0(n;t)$ in \eqref{4.2} and using \eqref{4.1a} we get the 
following identity
\begin{align}
&\sum_{i=1}^N\Bigg(\frac{w_{k+1,i}(t)}{n-x_i(t)}
+\frac{1}{n-x_i(t)} \frac{\pd w_{k,i}(t)}{\pd t_1}
+\frac{w_{k,i}(t)}{(n-x_i(t))^2}\frac{\pd x_i(t)}{\pd t_1}\Bigg)
                                                            \nonumber\\
&\quad =\sum_{i=1}^N\Bigg(\frac{w_{k+1,i}(t)}{n+1-x_i(t)}
+\frac{w_{k,i}(t)}{n+1-x_i(t)}
-\frac{w_{k,i}(t)}{n-x_i(t)}\Bigg)                          \label{4.4}\\
&\qquad + 
\Bigg(\sum_{i=1}^N\frac{1}{(n-x_i(t))(n+1-x_i(t))}
\frac{\pd x_i(t)}{\pd t_1}\Bigg)
\Bigg(\sum_{i=1}^N\frac{w_{k,i}(t)}{n-x_i(t)}\Bigg). 
\nonumber
\end{align}
Notice that \eqref{4.4} can be rewritten as a polynomial identity in $n$, 
which is true for every $n\in\Z$ and therefore, it will be true for every 
$n\in\C$. Computing the residue at $n=x_i(t)-1$ we obtain 
\begin{equation}\label{4.5}
-w_{k+1,i}(t)=w_{k,i}-\sum_{j=1}^{N}\frac{w_{k,j}(t)}{x_i(t)-x_j(t)-1}
\frac{\pd x_i(t)}{\pd t_1}.
\end{equation}
If we denote $\vw_k(t)=(w_{k,1}(t),w_{k,2}(t),\dots,w_{k,N}(t))^t$, 
$\ve=(1,1,\dots,1)^t$, then the last formula can be rewritten in vector 
notations as $\vw_{k+1}(t)=(-Y)\vw_k(t)$, where $Y$ is the matrix defined in 
\thref{th3.1}.
Thus we see that
\begin{equation}                                             \label{4.6}
\vw_k(t)=(-Y)^{k-1}\frac{\pd X}{\pd t_1}\ve, 
\end{equation}
where $X$ is the diagonal matrix given in equation \eqref{3.5a}.

Computing also the residue of equation \eqref{4.4} at $n=x_i(t)$ we obtain
\begin{align*}
&w_{k+1,i}(t)+\frac{\pd w_{k,i}(t)}{\pd t_1}=-w_{k,i}(t) 
+\frac{\pd x_i(t)}{\pd t_1}
\sum_{\begin{subarray}{c}j=1 \\j\neq i\end{subarray}}^N
\frac{w_{k,j}(t)}{x_i(t)-x_j(t)}\\
&\quad+w_{k,i}(t)\Bigg(-\frac{\pd x_i(t)}{\pd t_1}
+\sum_{\begin{subarray}{c}j=1 \\j\neq i\end{subarray}}^N
\frac{1}{(x_i(t)-x_j(t))(x_i(t)+1-x_j(t))}
\frac{\pd x_j(t)}{\pd t_1}\Bigg).
\end{align*}
Using the last identity and \eqref{4.5} we can eliminate $w_{k+1,i}(t)$. For 
$k=1$ this leads to the following second-order differential equation for 
$x_i(t)$
\begin{equation}\label{4.7}
\begin{split}
&\frac{\pd^2 x_i(t)}{\pd t_{1}^2}=-2\frac{\pd x_i(t)}{\pd t_1} \\
&\quad \times\sum_{\begin{subarray}{c}j=1 \\j\neq i\end{subarray}}^N
\frac{1}{(x_i(t)-x_j(t))(x_i(t)-x_j(t)+1)(x_i(t)-x_j(t)-1)}
\frac{\pd x_j(t)}{\pd t_1},
\end{split}
\end{equation}
which will be needed later.

Similarly, if we work with the adjoint wave function $w^*(n;t,z)$ we 
can show that it satisfies the following equation
\begin{equation*}
\frac{\pd w^*(n;t,z)}{\pd t_1}=(\nabla-a_0(n-1;t))w^*(n;t,z).
\end{equation*}
Denoting $\vw^*_k(t)=(w^*_{k,1}(t),w^*_{k,2}(t),\dots,w^*_{k,N}(t))^t$ we 
obtain as above that 
\begin{equation}                                             \label{4.8}
\vw^*_k(t)=-\frac{\pd X}{\pd t_1}(-Y^t)^{k-1}\ve.
\end{equation}
Next, we use \eqref{2.3}. From equations \eqref{2.2} and \eqref{2.4} we deduce 
that
\begin{equation*}
L^k=W(n;t)\De^k W(n;t)^{-1}
=\sum_{i,j=0}^{\infty}w_i(n;t)\De^{k-i-j}\cdot w^*_j(n+1;t),
\end{equation*}
where $w_0(n;t)=w^*_0(n;t)=1$. This shows that 
\begin{equation*}
(L^k)_-=\sum_{j=0}^{k+1}w_{k+1-j}(n;t)w^*_{j}(n;t)\De^{-1}+O(\De^{-2}).
\end{equation*}
On the other hand 
\begin{equation*}
W(n;t)=1+\Bigg(\sum_{i=1}^N\frac{1}{n-x_i(t)}\frac{\pd x_i(t)}{\pd t_1}\Bigg)
\De^{-1}+O(\De^{-2}).
\end{equation*}
Plugging the last two formulas in \eqref{2.3}, and 
equating the coefficients of $\De^{-1}$ on both sides we get
\begin{align*}
&\sum_{i=1}^N\Bigg(\frac{1}{n-x_i(t)}\frac{\pd^2 x_i(t)}{\pd t_1\pd t_k}
+\frac{1}{(n-x_i(t))^2}\frac{\pd x_i(t)}{\pd t_1}\frac{\pd x_i(t)}{\pd t_k}
\Bigg) \\
&\quad =-\sum_{j=0}^{k+1}w_{k+1-j}(n;t)w^*_{j}(n;t).
\end{align*}
The last equality holds for every $n\in\Z$ and therefore it must hold for 
every $n\in \C$. Comparing the coefficients of $(n-x_i(t))^{-2}$ gives
\begin{equation*}
\frac{\pd x_i(t)}{\pd t_1}\frac{\pd x_i(t)}{\pd t_k} =
-\sum_{j=1}^{k}w_{k+1-j,i}(t)w^*_{j,i}(t).
\end{equation*}
Let us denote by $I_i$ the elementary $N\times N$ matrix having 1 at entry 
$(i,i)$ and $0$ everywhere else. Using the last identity, \eqref{4.6} and 
\eqref{4.8} we obtain 
\begin{align*}
&\frac{\pd x_i(t)}{\pd t_1}\frac{\pd x_i(t)}{\pd t_k} =
-\sum_{j=1}^{k}\vw_{k+1-j}^t(t)I_i\vw^*_{j}(t)\\
&\qquad =(-1)^{k+1}\sum_{j=1}^k \ve^{\,t} Y^{j-1}\frac{\pd X}{\pd t_1} 
I_i Y^{k-j}\frac{\pd X}{\pd t_1}\ve .
\end{align*}
Notice that $\pd X/\pd t_1 I_i=\pd x_i(t)/\pd t_1 I_i$ and therefore, we can 
cancel $\pd x_i(t)/\pd t_1$ and the last formula reduces to 
\begin{equation*}
\frac{\pd x_i(t)}{\pd t_k}=(-1)^{k+1}\sum_{j=1}^k \ve^{\,t} Y^{j-1} 
I_i Y^{k-j}\frac{\pd X}{\pd t_1}\ve 
\end{equation*}
For every $N\times N$ matrix $A$ we have 
$\ve^{\,t}A\ve=\tr(A\ve\ve^{\,t})$, and thus we can rewrite the last formula 
for $\pd x_i(t)/\pd t_k$ as follows
\begin{equation*}
\frac{\pd x_i(t)}{\pd t_k}=(-1)^{k+1}\tr\Bigg(\sum_{j=1}^k Y^{j-1} 
I_i Y^{k-j}\frac{\pd X}{\pd t_1}\ve\ve^{\,t}\Bigg).
\end{equation*}
From the definitions of matrices $X$ and $Y$ it is easy to see that 
$\frac{\pd X}{\pd t_1}\ve\ve^{\,t}=-(XY-YX-Y+I)$. Making this substitution and 
using the fact that $\tr(AB)=\tr(BA)$ we get
\begin{align*}
&\frac{\pd x_i(t)}{\pd t_k}=(-1)^{k}\tr\Bigg(\sum_{j=1}^k Y^{j-1} 
I_i Y^{k-j}(XY-YX-Y+I)\Bigg) \\
&\quad =(-1)^{k}\tr\sum_{j=1}^k\Bigg(I_iY^{k-j}XY^{j}-I_iY^{k-j+1}XY^{j-1}
-I_iY^{k}+I_iY^{k-1}\Bigg)\\
&\quad =(-1)^{k}\tr\big(I_i(XY^k-Y^kX-kY^k+kY^{k-1})\big).
\end{align*}
However, $I_iX=XI_i$ and therefore $\tr(I_iXY^{k})=\tr(I_iY^{k}X)$. This 
shows that
\begin{equation}                                           \label{4.9}
\frac{\pd x_i(t)}{\pd t_k}=k(-1)^{k}\,\tr\big((I_i-I_iY)Y^{k-1}\big).
\end{equation}
On the other hand it is easy to see that 
\begin{equation*}
\frac{\pd Y}{\pd y_i}=I_i-I_iY.
\end{equation*}
Thus
\begin{align*}
&\frac{\pd }{\pd y_i}\tr(Y^k)=\tr\Bigg(\sum_{j=1}^kY^{j-1}
\frac{\pd Y}{\pd y_i}Y^{k-j}\Bigg)
=\tr\Bigg(\sum_{j=1}^k
\frac{\pd Y}{\pd y_i}Y^{k-1}\Bigg)\\
&\quad =k\,\tr\Bigg(\frac{\pd Y}{\pd y_i}Y^{k-1}\Bigg)
=k\,\tr\big((I_i-I_iY)Y^{k-1}\big).
\end{align*}
The last formula combined with \eqref{4.9} gives the first equation in 
\eqref{3.3}. In order to prove that the second equation holds, we first 
notice that \eqref{4.7} is equivalent to the Lax equation 
\begin{equation}\label{4.10}
\frac{\pd Y}{\pd t_1}=[Y,M],
\end{equation}
where $M$ is an $N\times N$ matrix with entries
\begin{equation*}
\begin{split}
M_{i,j}& =-\frac{1}{x_i(t)-x_j(t)}\frac{\pd x_i(t)}{\pd t_1} \qquad
\text{ for }i\neq j\\
M_{i,i}& =\sum_{k=1}^N\frac{1}{x_i(t)-x_k(t)+1}\frac{\pd x_k(t)}{\pd t_1}
-\sum_{\begin{subarray}{c}k=1 \\k\neq i\end{subarray}}^N
\frac{1}{x_i(t)-x_k(t)}\frac{\pd x_k(t)}{\pd t_1}.
\end{split}
\end{equation*}
Differentiating \eqref{3.1} with respect to $t_k$ we get
\begin{equation}\label{4.11}
\frac{\pd y_i(t)}{\pd t_k} = 
-\frac{\pd }{\pd t_k} \log\Bigg(\frac{\pd x_i(t)}{\pd t_1}\Bigg)
+\frac{\pd}{\pd t_k}\log\Bigg(
\prod_{\begin{subarray}{c}s=1 \\s\neq i\end{subarray}}^N
\frac{x_i(t)-x_s(t)+1}{x_i(t)-x_s(t)}\Bigg).
\end{equation}
The derivative in the second term on the right-hand side of \eqref{4.11} can 
be evaluated using \eqref{4.9}
\begin{align*}
&\frac{\pd}{\pd t_k}\log\Bigg(
\prod_{\begin{subarray}{c}s=1 \\s\neq i\end{subarray}}^N
\frac{x_i(t)-x_s(t)+1}{x_i(t)-x_s(t)}\Bigg)\\ 
&\qquad =
k(-1)^k\tr\Bigg(\sum_{j=1}^N\Bigg(\frac{\pd}{\pd x_j}\log
\prod_{\begin{subarray}{c}s=1 \\s\neq i\end{subarray}}^N
\frac{x_i-x_s+1}{x_i-x_s}\Bigg)I_j(I-Y)Y^{k-1}\Bigg).
\end{align*}
For the first term we use both \eqref{4.9} and \eqref{4.10}:
\begin{align*}
&\frac{\pd }{\pd t_k} \log\Bigg(\frac{\pd x_i(t)}{\pd t_1}\Bigg) \\
&\quad = k(-1)^k\Big(\frac{\pd x_i(t)}{\pd t_1}\Big)^{-1}\frac{\pd }{\pd t_1}
\tr(I_iY^{k-1}-I_iY^{k})\\
&\quad = k(-1)^k\Big(\frac{\pd x_i(t)}{\pd t_1}\Big)^{-1}
\tr\Bigg(\sum_{j=1}^{k-1}I_iY^{j-1}[Y,M]Y^{k-1-j}
-\sum_{j=1}^{k}I_iY^{j-1}[Y,M]Y^{k-j}\Bigg)\\
&\quad = k(-1)^k\Big(\frac{\pd x_i(t)}{\pd t_1}\Big)^{-1}
\tr\Big(\left(MI_i(I-Y)-(I-Y)I_iM\right)Y^{k-1}\Big).
\end{align*}
To simplify the formulas, let us denote $\Yh=I-Y$ and 
$(\Mh)_{i,j}=(1-\delta_{i,j})M_{i,j}$ (i.e. $\Mh$ is the matrix obtained from 
$M$ by replacing the diagonal entries with zeros). Then, the last 
two formulas combined with \eqref{4.11} show that 
\begin{equation}\label{4.12}
\frac{\pd y_i(t)}{\pd t_k}=k(-1)^{k+1} \tr(BY^{k-1}),
\end{equation}
where 
\begin{equation*}
B=\Big(\frac{\pd x_i(t)}{\pd t_1}\Big)^{-1}(\Mh I_i\Yh-\Yh I_i\Mh)
-\sum_{j=1}^N\Bigg(\frac{\pd}{\pd x_j}\log
\prod_{\begin{subarray}{c}s=1 \\s\neq i\end{subarray}}^N
\frac{x_i-x_s+1}{x_i-x_s}\Bigg)I_j\Yh.
\end{equation*}
A straightforward computation now shows that
\begin{equation*}
B=\frac{\pd Y}{\pd x_i}+\Bigg[
\sum_{j=1}^N\frac{1}{x_j-x_i+1}I_j
-\sum_{\begin{subarray}{c}j=1 \\j\neq i\end{subarray}}^N\frac{1}{x_j-x_i}I_j
,Y\Bigg],
\end{equation*}
which combined with \eqref{4.12} gives
\begin{equation*}
\frac{\pd y_i(t)}{\pd t_k}
=k(-1)^{k+1} \tr\Bigg(\frac{\pd Y}{\pd x_i}Y^{k-1}\Bigg)=
(-1)^{k+1}\frac{\pd }{\pd x_i}\tr(Y^{k}),
\end{equation*}
completing the proof of \eqref{3.3}.

Conversely, assume now that \eqref{3.3} holds. Let us consider 
$\{x_i(t),y_i(t)\}_{i=1}^N$ and the corresponding matrices $X$ and $Y$
 at the initial time $t_1=t_2=\cdots=0$ and let us denote 
$x_i^0=x_i(0)$, $y_i^0=y_i(0)$, $X^0=X|_{t=0}$, $Y^0=Y|_{t=0}$. 
Notice that 
\begin{equation*}
\rank(X^0Y^0-Y^0X^0+I-Y^0)=1.
\end{equation*}
Using the Cauchy determinant formula we see that 
\begin{equation*}
\det(I-Y^0)= e^{-\sum_{i=1}^{N}y_i^0} \neq 0.
\end{equation*}
Thus, if we denote $\Xt^0=X^0(I-Y^0)^{-1}$ we have
\begin{equation*}
\rank([\Xt^0,Y^0]+I)=1.
\end{equation*}
Therefore the pair $(\Xt^0,Y^0)$ defines a plane in Wilson's adelic 
Grassmannian $\Grad$, see \cite{W2}. The corresponding $\tau$-function 
can be computed by Shiota's formula
\begin{equation*}
\taut^0(t)=\det\Big(-\Xt^0+\sum_{j=1}^{\infty}jt_j(-Y^0)^{j-1}\Big),
\end{equation*}
see \cite[Corollary 1, p.~5845]{Sh}. 
Applying \cite[Theorem 2.4, p.~290]{HI} we deduce that 
$\taut^0(t_1+n,t_2-n/2,t_3+n/3,\dots)$ is 
a $\tau$-function for the discrete KP hierarchy \eqref{2.1}. Multiplying 
by the nonzero constant factor $\det(I-Y^0)$ we see that
\begin{equation*}
\begin{split}
\taut(n;t)
&=\det(I-Y^0)\taut^0\big(t_1+n,t_2-\frac{n}{2},t_3+\frac{n}{3},\dots\big)\\
&=\det\Big(nI-X^0+\sum_{j=1}^{\infty}jt_j(I-Y^0)(-Y^0)^{j-1}\Big)
\end{split}
\end{equation*}
is a $\tau$-function for the discrete KP hierarchy. Clearly, $\taut(n;t)$ 
is a monic polynomial in $n$, and therefore, by the first part of theorem, its 
roots $\xt_i(t)$ and the corresponding $\yt_i(t)$ will 
satisfy the Hamiltonian systems \eqref{3.3}. 
To complete the proof we show that 
$\tau(n;t)$ given by \eqref{3.2} coincides with $\taut(n;t)$ defined 
above. Since the roots of $\tau(n;t)$ and $\taut(n;t)$
satisfy the same systems \eqref{3.3}, 
it is enough to show that $x_j^0=\xt_j(0)$ and $y_j^0=\yt_j(0)$.
This follows easily from the explicit formula for $\taut(n;t)$:
\begin{align*}
&\taut(n;t_1,0,0\dots)=\det(nI-X^0+t_1(I-Y^0))\\
&\qquad=\prod_{j=1}^N\Bigg(n-x_j^0+t_1e^{-y_j^0}
\prod_{\begin{subarray}{c}s=1 \\s\neq j\end{subarray}}^N
\frac{x_j^0-x_s^0+1}{x_j^0-x_s^0}\Bigg)+O(t_1^2).
\end{align*}

\begin{Remark} It would be interesting to see if one can use the explicit 
formulas for $\tau$-functions of KP hierarchy in terms of matrices satisfying 
rank one conditions \cite{GK} and the construction of $\tau$-functions for 
$q$-KP hierarchy from classical ones to extend the above proof and to 
show that every solution of the $q$-deformed Calogero-Moser hierarchy 
described in \cite[Theorem 6.1]{I2} leads to a $\tau$-function for $q$-KP. 
This would give a one to one correspondence between rational solutions to the 
$q$-KP hierarchy (which also parametrize rank one solutions to a bispectral 
problem for $q$-difference operators) and $q$-deformed Calogero-Moser 
type systems.
\end{Remark}

\bigskip\noindent
{\bf Acknowledgments.} I thank a referee for his suggestions to
improve an earlier version of the paper.



\begin{thebibliography}{xx}

\bibitem{A} M.~Adler,  {\em Some finite dimensional integrable systems and 
their scattering behavior}, Comm. Math. Phys. {\bf 55} (1977), no. 3, 
195--230.

\bibitem{AM} M.~Adler and J.~Moser, {\em On a class of polynomials connected 
with the Korteweg-de Vries equation},  Comm. Math. Phys.  {\bf 61}  (1978), 
no. 1, 1--30.

\bibitem{AvM} M.~Adler and P.~van~Moerbeke, {\em Vertex operator solutions to 
the discrete KP-hierarchy}, Comm. Math. Phys. {\bf 203} (1999), no. 1, 
185--210.

\bibitem{AMcKM} H.~Airault, H.~P.~McKean and J.~Moser, {\em Rational and 
elliptic solutions of the Korteweg-de Vries equation and a related many-body 
problem}, Comm. Pure Appl. Math. {\bf 30} (1977), no. 1, 95--148.


\bibitem{BW} Yu.~Berest and G.~Wilson, {\em Ideal classes of the Weyl algebra 
and noncommutative projective geometry}, (with an appendix by 
M.~Van~den~Bergh), Internat. Math. Res. Notices 
{\bf 2002} (2002), no. 26, 1347--1396. 

\bibitem{C} F.~Calogero, {\em Solution of the one-dimensional $N$-body 
problems with quadratic and/or inversely quadratic pair potentials}, 
J. Math. Phys. {\bf 12} (1971), no. 3, 419--436.

\bibitem{Di} L.~A.~Dickey, {\em Soliton Equations and Hamiltonian 
Systems}, Advanced Series in Mathematical Physics, {\bf 26}, World Scientific 
Publishing Co., Inc., River Edge, NJ, 2003.

\bibitem{DG} J.~J.~Duistermaat and F.~A.~Gr\"unbaum, {\em Differential 
equations in the spectral parameter}, Comm. Math. Phys. {\bf 103} (1986), 
no. 2, 177--240.

\bibitem{GK} M.~Gekhtman and A.~Kasman, {\em Integrable systems and conditions 
for rectangular matrices to be of rank one}, Theor. Math. Phys. {\bf 133} 
(2002), no. 2, 1498--1503.

\bibitem{Gr} F.~A.~Gr\"unbaum, The limited angle reconstruction problem, in:  
{\em Computed tomography (Cincinnati, Ohio, 1982)},  pp. 43--61, 
Proc. Sympos. Appl. Math., {\bf 27}, Amer. Math. Soc., Providence, R.I., 1982.

\bibitem{H}  L.~Haine, The Bochner-Krall problem: some new perspectives, in: 
{\em Special functions 2000: current perspective and future directions (Tempe, 
AZ)},  141--178, NATO Sci. Ser. II Math. Phys. Chem., {\bf 30}, Kluwer Acad. 
Publ., Dordrecht, 2001.

\bibitem{HI} L. Haine and P. Iliev, {\em Commutative rings of difference 
operators and an adelic flag manifold}, Internat. Math. Res. Notices 
{\bf 2000} (2000), no. 6, 281--323.

\bibitem{Hor} E.~Horozov, {\em Calogero-Moser spaces and an adelic 
$W$-algebra}, Ann. Inst. Fourier (Grenoble)  {\bf 55}  (2005),  no. 6, 
2069--2090. 
 
\bibitem{I2} P.~Iliev, {\em $q$-KP hierarchy, bispectrality and Calogero-Moser 
systems}, J. Geom. Phys. {\bf 35} (2000),  no. 2-3, 157--182. 

\bibitem{KKS} D.~Kazhdan, B.~Kostant and S.~Sternberg, {\em Hamiltonian 
group actions and dynamical systems of Calogero type}, Comm. Pure Appl. 
Math. {\bf 31} (1978), no. 4, 481-507.

\bibitem{K} I.~M.~Krichever, {\em Rational solutions of the 
Kadomtsev-Petviashvili equation and integrable systems of $n$ particles on a 
line}, Funct. Anal. Appl. {\bf 12} (1978), 59-61.

\bibitem{KZ} I.~M.~Krichever and A.~Zabrodin, {\em Spin generalization of the 
Ruijsenaars-Schneider model, the nonabelian two-dimensionalized Toda lattice, 
and representations of the Sklyanin algebra}, 
Russian Math. Surveys {\bf 50} (1995), no. 6, 1101--1150.


\bibitem{M} J.~Moser, {\em Three integrable Hamiltonian systems connected with 
isospectral deformations}, Adv. Math. {\bf 16} (1975), no. 2, 197--220. 

\bibitem{R} S.~N.~M.~Ruijsenaars, {\em Integrable particle systems vs 
solutions to the KP and $2$D Toda equations}, 
Ann. Physics {\bf 256} (1997), no. 2, 226--301.

\bibitem{RS} S.~N.~M.~Ruijsenaars and H.~Schneider, {\em A new class of 
integrable systems and its relation to solitons}, 
Ann. Physics {\bf 170} (1986), no. 2, 370--405.

\bibitem{SS} M.~Sato and Y.~Sato, {\em Soliton equations as dynamical systems 
on infinite dimensional Grassmann manifolds}, Lect. Notes Num. Appl. Anal. 
{\bf 5 } (1982), 259--271.

\bibitem{Sh} T.~Shiota, {\em Calogero-Moser hierarchy and KP hierarchy}, 
J. Math. Phys. {\bf 35} (1994), no. 11, 5844-5849.

\bibitem{UT}  K.~Ueno and K.~Takasaki, {Toda lattice hierarchy}, in: 
{\em Group representations and systems of differential equations 
(Tokyo, 1982)}, 1--95, Adv. Stud. Pure Math., {\bf 4}, North-Holland, 
Amsterdam, 1984. 

\bibitem{vD} J.~F.~van~Diejen, {\em The dynamics of zeros of the solitonic 
Baker-Akhiezer function for the Toda chain}, 
Internat. Math. Res. Notices {\bf 2000} (2000), no. 5, 253--270.

\bibitem{vDP} J.~F.~van~Diejen and  H.~Puschmann, 
{\em Reflectionless Schr\"odinger operators, the dynamics of zeros, and the 
solitonic Sato formula}, Duke Math. J. {\bf 104} (2000), no. 2, 269--318.

\bibitem{W1} G.~Wilson, {\em Bispectral commutative ordinary differential 
operators}, J. Reine Angew. Math. {\bf 442} (1993), 177--204.

\bibitem{W2} G.~Wilson, {\em Collisions of Calogero-Moser particles and an 
adelic Grassmannian}, (with an appendix by I. G. Macdonald),  
Invent. Math. {\bf 133} (1998), no. 1, 1--41.



\end{thebibliography}
\end{document}